\documentstyle[floats,aps]{revtex}
\begin{document}
\input epsf
\hsize=6.5truein
\hoffset=0.0truein
\vsize=9.0truein
\voffset=0.6truein
\hfuzz=0.1pt
\vfuzz=0.1pt
\parskip=\medskipamount
\overfullrule=0pt

  \font\twelvemib=cmmib10 scaled 1200
  \font\elevenmib=cmmib10 scaled 1095
  \font\tenmib=cmmib10
  \font\eightmib=cmmib10 scaled 800
  \font\sixmib=cmmib10 scaled 667
  \skewchar\elevenmib='177
  \newfam\mibfam
  \def\mib{\fam\mibfam\tenmib}
  \textfont\mibfam=\tenmib
  \scriptfont\mibfam=\eightmib
  \scriptscriptfont\mibfam=\sixmib

\draft

\twocolumn[\hsize\textwidth\columnwidth\hsize\csname  
@twocolumnfalse\endcsname

\title{Kinky Behavior in Josephson Junctions}
\author{A. V. Rozhkov and Daniel P. Arovas}
\address{Department of Physics, University of California at San Diego,
La Jolla CA 92093}

\date{\today}

\maketitle

\begin{abstract}
We analyze nonperturbatively the behavior of a Josephson junction
in which two BCS superconductors are coupled through an Anderson impurity.
We recover earlier perturbative results which found that a $\delta=\pi$ phase
difference is preferred when the impurity is singly occupied and the
on-site Coulomb interaction is large.  We find a novel intermediate
phase in which one of $\delta=0$ and $\delta=\pi$ is stable while the
other is metastable, with the energy $E(\delta)$ having a kink somewhere
in between.  As a consequence of the kink, the $I-V$ characteristics
of the junction are modified at low voltages.
\end{abstract}

\vskip2pc]

\narrowtext

\section{Introduction}

Two identical superconductors separated by a barrier will generate a
Josephson current $J(\delta)=J_ {\rm c}\sin(\delta)$, where $\delta$ is
the phase difference between the superconductors.  $J_ {\rm c}$ is
given by the Ambegaokar-Baratoff formula $J_ {\rm c}=\pi\Delta/2eR$,
where $\Delta$ is the superconducting gap and $R$ is the normal state
resistance of the barrier.  This result is derived perturbatively,
assuming some (spin-conserving) tunneling amplitude $t$ for electrons
to move across the barrier region;
the conductance $G=1/R$ is then proportional to $|t|^2$ when $t$ is small.
In the 1960's, Kulik \cite{kulik} showed that tunneling processes which
do not conserve spin act to decrease $J_ {\rm c}$ and potentially drive it
negative, so that $J_ {\rm c}\propto |t|^2 - |t ^{\vphantom{\dagger}}_{\rm sf}|^2$, where
$t ^{\vphantom{\dagger}}_{\rm sf}$ is the amplitude for spin flip tunneling across the barrier.

This result was generalized to the case of tunneling through a
dynamical (Kondo or Anderson) impurity by Shiba and Soda \cite{shiba}
and by Glazman and Matveev \cite{glma}.  The essence of negative Josephson
coupling was also elucidated by Spivak and Kivelson \cite{spki}, who
considered an interacting barrier region comprised of a single Anderson
impurity.  Assuming the local Hubbard $U$ is large and the impurity is
singly occupied, they showed that the fourth order process
\begin{center}
\begin{tabular}{||c|c|c|}
\hline
SC \#1  &  IMPURITY  & SC \#2 \\ \hline\hline
$ \uparrow \downarrow$ & $ \uparrow$ & -- \\ \hline
$ \uparrow \downarrow$ & -- & $ \uparrow$ \\ \hline
$ \uparrow$ & $ \downarrow$ & $ \uparrow$ \\ \hline
$ \uparrow$ & --  & $ \downarrow \uparrow$ \\ \hline
-- & $ \uparrow$ & $ \downarrow \uparrow$ \\ \hline
\hline
\end{tabular}
\end{center}
reverses the order of spins in the Cooper pair, and hence leads to
a negative Josephson coupling.  For an Anderson impurity with site
energy $  \varepsilon ^{\vphantom{\dagger}}_0$, the condition for single occupancy (for infinitesimal
hopping) is $U > -  \varepsilon ^{\vphantom{\dagger}}_0 > 0$.

When $J_ {\rm c}<0$, the most favorable configuration for the junction is
one in which the phase difference is $\delta=\pi$.  In a ring with a
single such $\pi$-junction, time reversal is broken and there is a
trapped flux of $\pm hc/4e$ \cite{bula}.  In recent years, several experiments
experiments \cite{Harlingen} have focussed on the existence of $\pi$-junctions
in high $T_ {\rm c}$ superconductors and have been invoked as evidence for
the existence of a $d ^{\vphantom{\dagger}}_{x^2-y^2}$ wave order parameter therein.

In this paper, we are again concerned with simple $s$-wave superconductors
coupled via an Anderson impurity.  We rederive earlier results on
$\pi$-junctions using nonperturbative techniques, and we find a crossover
region, which separates $0$- and $\pi$-junction behavior.  This crossover
regime is characterized by a ground state energy $E(\delta)$ which has local
minima at both $\delta=0$ and $\delta=\pi$, and a kink at its maximum, 
occuring at an intermediate value of $\delta$.  We derive a phase diagram
as a function of $-  \varepsilon ^{\vphantom{\dagger}}_0/\Delta$, $U/\Delta$, and $\Gamma/\Delta$, where
$\Gamma$ is the bare impurity level width generated by virtual hopping onto
the superconductors.  We also discuss the $I(V)$ behavior of such a junction
within the resistively shunted junction (RSJ) model.

\section{Effective Action}

Consider two superconducting planes each of which is connected via electron
hopping to an Anderson impurity.  For electrons living in a fully
two-dimensional space, the field operator $\psi( {\mib r})$ may be
decomposed into partial waves \cite{Kondo,ChamFrad},
\begin{equation}
\psi ^\dagger({\mib r})=\sum_{l=-\infty}^\infty e^{il\phi}\int\limits_0^\infty\!
{dk\over 2\pi}\,\sqrt{k}\, J_l(kr)\,\psi ^\dagger_l(k)
\label{pwave}
\end{equation}
where $\{\psi ^{\vphantom{\dagger}}_{l}(k),\psi ^\dagger_{l'}(k')\}=2\pi\,
\delta ^{\vphantom{\dagger}}_{ll'}\,\delta(k-k')$ and where spin indices
have been suppressed.  Within each $l$ sector, the kinetic energy can
be linearized to $\varepsilon\simeq \hbar v_{\scriptscriptstyle{\rm F}} (k- k_{\scriptscriptstyle{\rm F}})$.  Extending the lower
limit of $k$ from $0$ to $-\infty$ is innocuous provided we are interested
in low energy properties which only involve states for which
$|k- k_{\scriptscriptstyle{\rm F}}|\ll  k_{\scriptscriptstyle{\rm F}}$.  Each $l$ sector, then, yields a single chiral
fermion branch on the real line (or two chiral fermions on a half line):
\begin{eqnarray}
\psi ^\dagger_l(x)&\equiv&\int\limits_{-\infty}^\infty\!\!{dk\over 2\pi}
\,e^{ikx}\,\psi ^\dagger_l(k)\nonumber\\
 {\cal T} ^{\vphantom{\dagger}}_l&=&\hbar v_{\scriptscriptstyle{\rm F}}\int\limits_{-\infty}^\infty\!\!\! dx\,\psi ^\dagger_l(x)
\left({1\over i}{ {\partial}\over {\partial} x}- k_{\scriptscriptstyle{\rm F}}\right)\psi ^{\vphantom{\dagger}}_l(x)\ .
\label{hlin}
\end{eqnarray}
Since $\psi({\bf 0})\simeq\sqrt{ k_{\scriptscriptstyle{\rm F}}}\,\psi ^{\vphantom{\dagger}}_{l=0}(0)$, so only the
$l=0$ channel is involved in connecting the two planes.

Modeling the superconductors with the BCS Hamiltonian, we make use of
\begin{eqnarray}
\int\!d^2\!r\, \psi ^\dagger_\uparrow({\mib r})\,
\psi ^\dagger_\downarrow({\mib r})&=&\int\limits_0^\infty
{dk\over 2\pi}\,\sum_l (-1)^l\,\psi ^\dagger_{l \uparrow}(k)\,\psi ^\dagger_{-l \downarrow}(k)
\nonumber\\
\simeq\sum_l(-1)^l&&\!\!\!\!\int\limits_0^\infty dx\,
\psi ^\dagger_{l \uparrow}(x)\,\psi ^\dagger_{-l \downarrow}(-x)\ ,
\end{eqnarray}
which is nonlocal in this representation.  It may be rendered local by means
of the canonical transformation
$ \psi ^\dagger_\uparrow(x)\rightarrow \psi ^\dagger_\uparrow(-x)$\ ,
which converts up spins into left movers.  We now integrate out the Fermi
fields at all points other than $x=0$;
this means that only the $l=0$ modes contribute.  Defining
$ \psi ^\dagger_\sigma(\tau)\equiv\psi ^\dagger_{l=0,\sigma}(x=0,\tau)$\ ,
we obtain the effective action for the point $x=0$,
\begin{equation}
S ^{\vphantom{\dagger}}_{\rm eff}=\sum_{\omega_m}  {\bar\psi} ^{\vphantom{\dagger}}_i(\omega_m)\,R ^{\vphantom{\dagger}}_{ij}(\omega_m)\,
\psi ^{\vphantom{\dagger}}_j(\omega_m)
\end{equation}
where
\begin{eqnarray}
\psi ^{\vphantom{\dagger}}_i(\omega_m)&=&
\pmatrix{ \psi ^{\vphantom{\dagger}}_\uparrow(\omega_m)\cr
\bar\psi ^{\vphantom{\dagger}}_\downarrow(-\omega_m)\cr}\nonumber\\
R ^{\vphantom{\dagger}}_{ij}(\omega_m)&=&
{2 v_{\scriptscriptstyle{\rm F}}\over \sqrt{\omega_m^2+\Delta^2}}\pmatrix{-i\omega_m & \Delta\,e^{i\delta}\cr
\Delta\,e^{-i\delta} & -i\omega_m\cr}
\label{seffa}
\end{eqnarray}
Here, $\omega_m=2\pi(m+ \frac{1}{2})T$ is a fermionic Matsubara frequency,
$\{ \psi ^{\vphantom{\dagger}}_\sigma, \bar\psi ^{\vphantom{\dagger}}_\sigma\}$ are Grassmann fields, and $\delta$
is the phase of the superconducting condensate ($\Delta$ is real).

\begin{figure} [!t]
\centering
\leavevmode
\epsfxsize=8cm
\epsfysize=8cm
\epsfbox[18 144 592 718] {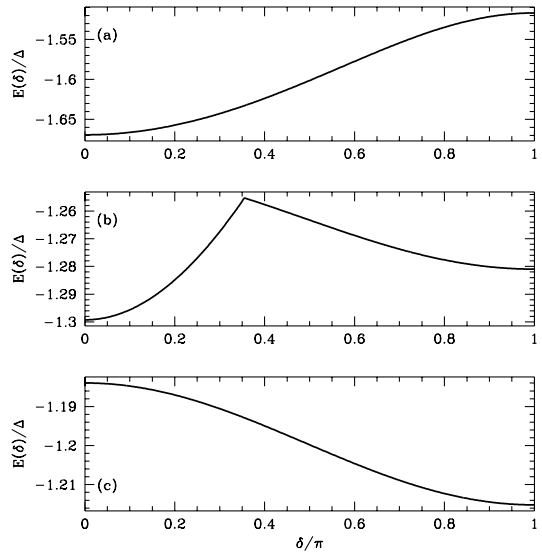}
\caption[]
{\label{erg} Ground state energy for junctions with $  \varepsilon ^{\vphantom{\dagger}}_0=-2$,
$\Gamma/\Delta=1$, and (a) $U=2.1$, (b) $U=2.6$, and (c) $U=2.9$.}
\end{figure}

\begin{figure} [!t]
\centering
\leavevmode
\epsfxsize=8cm
\epsfysize=8cm
\epsfbox[18 144 592 718] {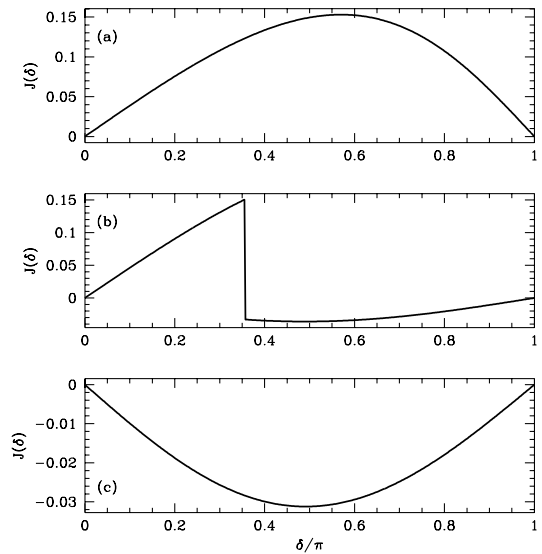}
\caption[]
{\label{cur} Josephson current (in units of $2e\Delta/\hbar$) for junctions with
$  \varepsilon ^{\vphantom{\dagger}}_0=-2$, $\Gamma/\Delta=1$, and (a) $U=2.1$, (b) $U=2.6$, and (c) $U=2.9$.}
\end{figure}

The superconductor is thus reduced the zero-dimensional action
of equation (\ref{seffa}).  The on-site interactions of
the Anderson impurity and its coupling to the superconductors
is described by the Hamiltonian\cite{foot1}
\begin{eqnarray}
 {\cal H} ^{\vphantom{\dagger}}_1&=&-{t\over\sqrt{ k_{\scriptscriptstyle{\rm F}}}}\sum_{\sigma}\left(  \psi ^\dagger_\sigma  c ^{\vphantom{\dagger}}_\sigma+  c ^\dagger_\sigma \psi ^{\vphantom{\dagger}}_\sigma\right) 
+  \varepsilon ^{\vphantom{\dagger}}_0\sum_\sigma  c ^\dagger_\sigma c ^{\vphantom{\dagger}}_\sigma\nonumber\\
&&\qquad +U  c ^\dagger_\uparrow  c ^\dagger_\downarrow  c ^{\vphantom{\dagger}}_\downarrow  c ^{\vphantom{\dagger}}_\uparrow\ .
\end{eqnarray}
When $t=0$, the ground state of the impurity is singly occupied provided
$U>-  \varepsilon ^{\vphantom{\dagger}}_0>0$.  We couple the Anderson impurity to two superconductors,
assumed identical in every respect except in their phase.  We may then
integrate out the $\{ \psi ^{\vphantom{\dagger}}_\sigma, \bar\psi ^{\vphantom{\dagger}}_\sigma\}$ fields on each superconductor and
decouple the interaction term via a Hubbard-Stratonovich (HS) transformation.
Neglecting temporal fluctuations in the HS field, we have
\begin{equation}
Z={1\over\sqrt{2\pi UT}} \int\limits_{-\infty}^\infty\!d\gamma\,
\exp\left( -{\gamma^2\over 2UT}-{ {\tilde \varepsilon} ^{\vphantom{\dagger}}_0\over T}\right) \,\prod_{\omega_m}
{\rm det}\,M(\omega_m)\ ,
\label{res}
\end{equation}
where $M(\omega_m)$ is the matrix
\begin{displaymath}
\pmatrix{-i\omega_m(1+\alpha_m)+ {\tilde \varepsilon} ^{\vphantom{\dagger}}_0+\gamma&
-\alpha_m\,\Delta\,\cos(\delta/2)\cr
-\alpha_m\,\Delta\,\cos(\delta/2)&
-i\omega_m(1+\alpha_m)- {\tilde \varepsilon} ^{\vphantom{\dagger}}_0+\gamma\cr}\ ,
\end{displaymath}
where $\alpha_m\equiv\alpha(\omega_m)\equiv\Gamma/ \sqrt{\omega_m^2+\Delta^2}$ and
$\Gamma=t^2/ k_{\scriptscriptstyle{\rm F}} v_{\scriptscriptstyle{\rm F}}$ is the (bare) impurity level width.
$\delta=\delta_1-\delta_2$ is the phase difference between the
two superconductors, and $ {\tilde \varepsilon} ^{\vphantom{\dagger}}_0=  \varepsilon ^{\vphantom{\dagger}}_0+ \frac{1}{2} U$.
We now make an approximation by performing the integral in (\ref{res})
using the method of steepest descents.  That is, we compute the free energy
\begin{eqnarray}
F&=&-T\sum_{\omega_m}\ln\Big([\omega_m(1+\alpha_m)+i\gamma]^2+ {\tilde \varepsilon}_0^2
\nonumber\\
&&\quad+[\alpha_m\,\Delta\,\cos( \frac{1}{2}\delta)]^2\Big)
+{\gamma^2\over 2U}+ {\tilde \varepsilon} ^{\vphantom{\dagger}}_0\ ,
\label{free}
\end{eqnarray}
and minimize over $\gamma$.  We remark that this procedure is exact in
the large-$N$ limit of an Sp$(2N)$ model in which, for each
$ \uparrow$ and $ \downarrow$ internal degree of freedom there are $N$ flavors
$a=1,\ldots,N$.  The on-site interaction term for this model is
\begin{eqnarray}
 {\cal H} ^{\vphantom{\dagger}}_{1,{\rm int}}&=&-{U\over 2N}\sum_{a,b}
(c ^\dagger_{a \uparrow}c ^{\vphantom{\dagger}}_{a \uparrow}-c ^\dagger_{a \downarrow}c ^{\vphantom{\dagger}}_{a \downarrow})
(c ^\dagger_{b \uparrow}c ^{\vphantom{\dagger}}_{b \uparrow}-c ^\dagger_{b \downarrow}c ^{\vphantom{\dagger}}_{b \downarrow})\nonumber\\
&&\qquad+{U\over 2N}\sum_{a,\sigma}c ^\dagger_{a\sigma}c ^{\vphantom{\dagger}}_{a\sigma}\ .
\end{eqnarray}
The free energy in (\ref{free}) is then the free energy per flavor,
and $ {\tilde \varepsilon} ^{\vphantom{\dagger}}_0=  \varepsilon ^{\vphantom{\dagger}}_0+(U/2N)$.

\begin{figure} [!t]
\centering
\leavevmode
\epsfxsize=8cm
\epsfysize=8cm
\epsfbox[18 144 592 718] {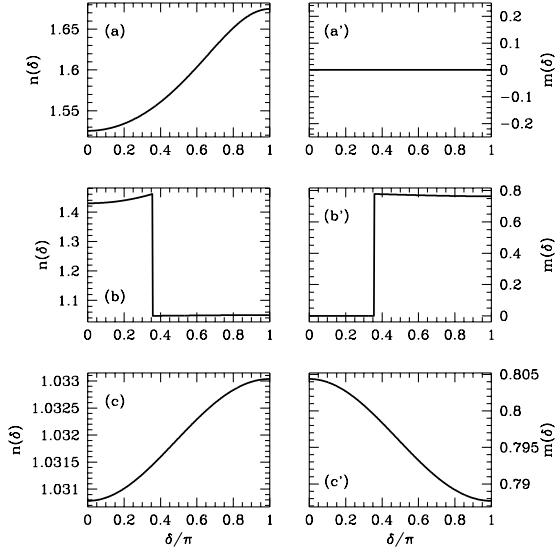}
\caption[]
{\label{nm} Occupation number $n$ and magnetization $m=\gamma^*/U$ for a
junction with $  \varepsilon ^{\vphantom{\dagger}}_0=-2$, $\Gamma/\Delta=1$, and (a,a') $U=2.1$, (b,b')
$U=2.6$, and (c,c') $U=2.9$.}
\end{figure}

\begin{figure} [!t]
\centering
\leavevmode
\epsfxsize=8cm
\epsfysize=8cm
\epsfbox[18 144 592 718] {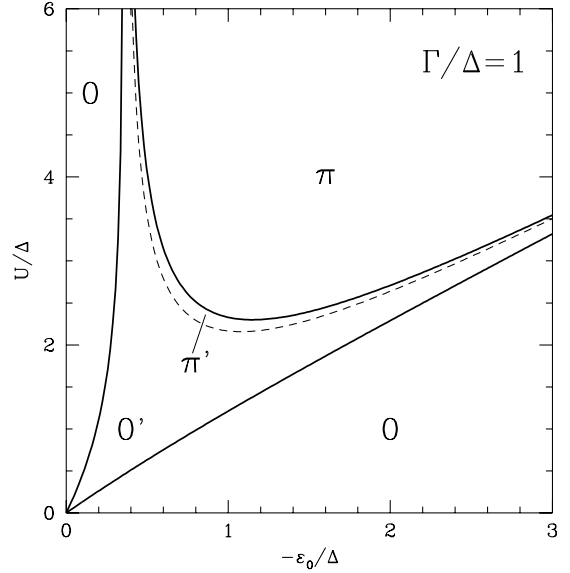}
\caption[]
{\label{pdiag} The phase diagram for $\Gamma/\Delta=1.0$.  In the ${\bf 0}$
phase (which includes all $  \varepsilon ^{\vphantom{\dagger}}_0>0$ as well), $\delta=0$ minimizes the energy of
the junction and $\delta=\pi$ is unstable.  In the $ {\mib\pi}$ phase, $\delta=\pi$
minimizes the junction energy and $\delta=0$ is unstable.  In the ${\bf 0}'$
phase, $\delta=0$ is the stable minimum while $\delta=\pi$ is metastable,
and {\it vice versa\/} for the $ {\mib\pi}'$ phase.}
\end{figure}

\begin{figure} [!t]
\centering
\leavevmode
\epsfxsize=8cm
\epsfysize=8cm
\epsfbox[18 144 592 718] {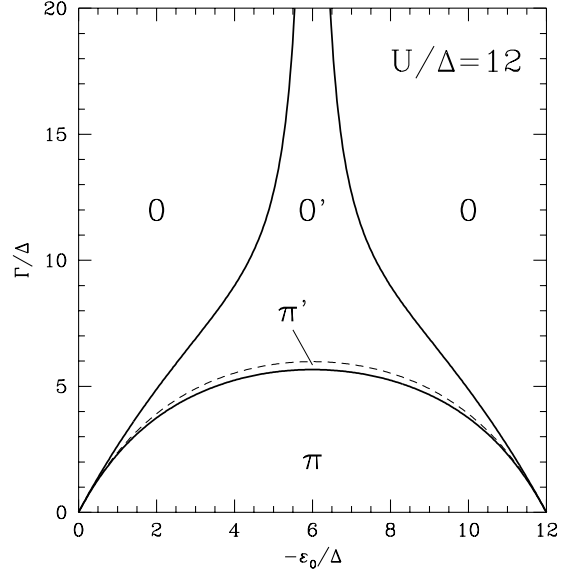}
\caption[]
{\label{pediag} The phase diagram for $U/\Delta=12$.  The phases ${\bf 0}$,
${\bf 0'}$, $ {\mib\pi}$, and $ {\mib\pi}'$ are as in FIG. \ref{pdiag}.}
\end{figure}

\section{Solution of the Model}

At $T=0$, the self-consistent equation for $\gamma$ is
\begin{eqnarray}
{ {\partial} E\over  {\partial} \gamma^2}&=&{1\over 2U}-
 \int\limits_0^\infty\!{d\omega\over\pi}\,{2\omega^2[1+\alpha(\omega)]^2
A(\omega)\over A^2(\omega)+4\gamma^2\omega^2[1+\alpha(\omega)]^2}=0\nonumber\\
A(\omega)&=&\omega^2[1+\alpha(\omega)]^2 -\gamma^2
+ {\tilde \varepsilon}_0^2+\alpha^2(\omega)\Delta^2\cos^2( \frac{1}{2}\delta)
\end{eqnarray}
Once the solution $\gamma=\gamma^*$ is obtained, we compute the current
$J= {\partial} E/ {\partial}\delta$ and the impurity occupancy $n= {\partial} E/ {\partial} {\tilde \varepsilon} ^{\vphantom{\dagger}}_0$,
\begin{eqnarray}
J&=&
 \int\limits_0^\infty\!{d\omega\over\pi}\,
{A(\omega)\,\alpha^2(\omega)\,\Delta^2\,\sin(\delta)\over
A^2(\omega)+4\gamma^2\omega^2[1+\alpha(\omega)]^2}\\
n&=&1- \int\limits_0^\infty\!{d\omega\over\pi}\,
{ {\tilde \varepsilon} ^{\vphantom{\dagger}}_0\,A(\omega)\over
A^2(\omega)+4\gamma^2\omega^2[1+\alpha(\omega)]^2}\ .
\end{eqnarray}
In the event that there is more than one solution for $\gamma^*$, we choose
the solution with the lowest energy.

In FIGs. \ref{erg}, \ref{cur}, and \ref{nm} we show the mean field solution
as we tune through a critical region of interaction strength $U$.  For low
$U$ the junction is unremarkable, with $J(\delta)\approx J_ {\rm c}\,\sin(\delta)$
and $J_ {\rm c}>0$.  (The deviation from a pure $\sin(\delta)$ behavior occurs
because $\Gamma/\Delta=1$
is not so small.)  The energy $E(\delta)$ exhibits a minimum at $\delta=0$,
and the curve $n(\delta)$ shows that the Anderson impurity is compressible
in that the occupancy responds to changes in $\delta$.  At high $U$,
the behavior is inverted: $J_ {\rm c}<0$ and $E(\delta)$ has a minimum at
$\delta=\pi$.  The occupancy of the Anderson impurity is effectively
pinned -- the impurity is incompressible.
This is the regime discussed by Spivak and Kivelson -- for $  \varepsilon ^{\vphantom{\dagger}}_0<0$ and
$U$ sufficiently large, the impurity maintains single occupancy
and the fourth order process (in $t$) which transfers a Cooper pair from
one superconductor to the other through the impurity reverses the order
of the up and down spins, thereby leading to a negative Josephson coupling.

The intermediate $U$ regime shown in the (b) panels of FIGS.
\ref{erg}, \ref{cur}, and \ref{nm} exhibits unusual
behavior.  Rather than the amplitude
of $J_ {\rm c}$ smoothly going through zero, we find that for $\delta \in
[0,\delta_ {\rm c}]$ the impurity is compressible and $J(\delta)>0$,
but further increase of $\delta$ results in an incompressible impurity
and a reversal of the Josephson current.  For the parameters given,
$E(\delta)$ has a global minimum at $\delta=0$ and a local minimum at
$\delta=\pi$; the relative positions of the minima will switch at a 
somewhat higher value of $U$.

A related result was derived in Ref. \cite{TanKash},
where the Josephson tunneling through a ferromagnet was studied.
The Josephson current was obtained as a sum of partial contributions due to
Andreev bound states in barrier region.  Certain bound states were found
to produce an anomalous phase dependence of the energy, similar to that
found here.

The phase diagram for $\Gamma/\Delta=1$ is shown in FIG. \ref{pdiag}.
In the phases marked ${\bf 0}$ and $ {\mib\pi}$, the only stable minima in
$E(\delta)$ lie at $\delta=0$ and $\delta=\pi$, respectively.  In the 
intermediate regime, both $\delta=0$ and $\delta=\pi$ are local minima;
the global minima then label the phases ${\bf 0}'$ and $ {\mib\pi}'$.
We have found that the qualitative shape of this phase diagram is valid
for all $\Gamma/\Delta$ we explored.  The phase diagram in the
$(-  \varepsilon ^{\vphantom{\dagger}}_0/\Delta,\Gamma/\Delta)$ plane is shown in FIG. \ref{pediag}.
The fact that the ${\bf 0'}$ phase persists in an ever-narrowing region
about $\varepsilon^{\vphantom{\dagger}}_0=-\frac{1}{2}\,U$ is an artifact of the large-$N$ model, which
is unable to describe the Kondo physics which sets in below temperatures
on the order of $T^{\vphantom{\dagger}}_{\rm K}\sim \Gamma\,\exp(-\pi
|\varepsilon^{\vphantom{\dagger}}_0|/2\Gamma)$ \cite{glma}.
The solid lines in FIGS. \ref{pdiag} and \ref{pediag},
all represent first order transitions, where all
quantities ($n$, $J$, $m$,  {\it etc.\/}) are discontinuous.

There is a symmetry in our model under $ {\tilde \varepsilon} ^{\vphantom{\dagger}}_0\to - {\tilde \varepsilon} ^{\vphantom{\dagger}}_0$,
 {\it i.e.\/}\ $-  \varepsilon ^{\vphantom{\dagger}}_0\to U+  \varepsilon ^{\vphantom{\dagger}}_0$, as is evident in FIG. \ref{pediag}.  The phase
boundaries in the lower left corner all start out linearly with the
same slope, which may be determined via evaluating the energy shifts
of the impurity states $|0\rangle$ and $|\uparrow\rangle$ doing perturbation
theory in the hopping $t$.  This line is determined to
$ {\cal O}(\Gamma)$ by $E_{|0\rangle}=E_{|\uparrow\rangle}$, which gives
\begin{equation}
-  \varepsilon ^{\vphantom{\dagger}}_0={\Gamma\over 4\pi} {U\over\sqrt{U^2-\Delta^2}}
\ln\left({U+\Delta+\sqrt{U^2-\Delta^2}\over U+\Delta-\sqrt{U^2-\Delta^2}}
\right)\ .
\end{equation}
This gives a slope $d\Gamma/d(-  \varepsilon ^{\vphantom{\dagger}}_0)$ which depends on $U/\Delta$.
Our large-$N$ theory tends to underestimate this slope.

\begin{figure} [!t]
\centering
\leavevmode
\epsfxsize=8cm
\epsfysize=8cm
\epsfbox[18 144 592 718] {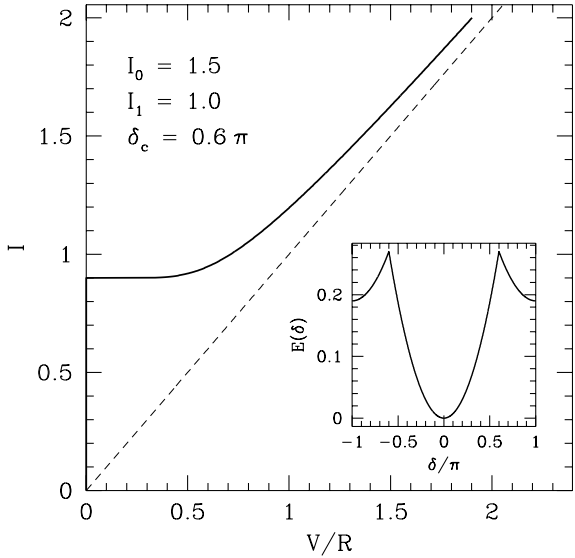}
\caption[]
{\label{iv} $I-V$ diagram for the simple analytical ${\bf 0}'$ junction model
(see text).  While qualitatively similar to the usual
$I=\sqrt{I_ {\rm c}^2+( {\bar V}/R)^2}$ behavior, here $I( {\bar V})$
flattens out much more rapidly as $V\to 0$.  The inset shows the 
energy $E(\delta)$.  This $I-V$ behavior also holds for
${ {\mib\pi}'}$-junctions.}
\end{figure}

How might a cusp in $E(\delta)$ manifest itself in an experiment on
a single junction?  Within the RSJ model, we have for $I>I_{\rm c}$ that
the average voltage $ {\bar V}$ across the junction is
\begin{equation}
 {\bar V} = R\bigg/\int\limits_0^{2\pi}\!{d\delta\over 2\pi}
\left[I-{2e\over\hbar}{ {\partial} E\over {\partial}\delta}\right]^{-1}\ .
\end{equation}
For a simple model of the ${\bf 0}'$ or ${ {\mib\pi}}'$ phase, we take
\begin{displaymath}
E(\delta)= \left\{ \begin{array}{ll}
(\hbar/4e)I_0\delta^2 &\mbox{if $0\le\delta\le\delta_ {\rm c}$}\\ &\\
E_\pi + (\hbar/4e) I_1(\delta-\pi)^2 &\mbox{if $\delta_ {\rm c}\le\delta\le
2\pi-\delta_ {\rm c}$}\\ &\\
(\hbar/4e)I_0 (\delta-2\pi)^2 &\mbox{if $2\pi-\delta_ {\rm c}\le\delta\le 2\pi$}
\end{array}\right.
\end{displaymath}
where $I_0$, $I_1$, $\delta_ {\rm c}$, and $E_\pi$ are related through continuity
of $E(\delta)$ at $\delta=\delta_ {\rm c}$, the location of the cusp.
The function $I( {\bar V})$ is shown in Fig. \ref{iv}.  The textbook
result, when $E=E_0(1-\cos\delta)$, is
$I=\sqrt{I_ {\rm c}^2+( {\bar V}/R)^2}$.  In our model, the cusp
results in a much flatter behavior as $ {\bar V}\to 0$, with
\begin{equation}
I( {\bar V})=I_ {\rm c}+\lambda\, I_ {\rm c}\,\exp(-V_0/ {\bar V})
\label{iveqn}
\end{equation}
where $I_ {\rm c}={\rm max}\big((\delta_ {\rm c}/\pi)\cdot I_0,\,(1-\delta_ {\rm c}/\pi)
\cdot I_1\big)$, and $\lambda$ and $V_0$ are constants.

It has been brought to our attention that two conventional Josephson
junctions in series will also yield a kink in $E(\delta)$ \cite{lgpc}.
If the junctions are identical, then at $T=0$ each is responsible for
half the total phase difference, and
$E(\delta)=2E_0\big(1-|\cos( \frac{1}{2}\delta)|\big)$.  However,
if each junction is resistively shunted, then individual $V(I)$
curves simply add, and one does not obtain the behavior in (\ref{iveqn}).

\section{Conclusions}
If single electron tunneling between superconductors flips the electron's
spin, then a Cooper pair is transferred with opposite sign.  This result,
as first shown by Kulik \cite{kulik}, can lead to a negative Josephson
coupling, in which case the ground state of the junction is one in which
there is a $\delta=\pi$ phase difference between the two superconductors.
This phenomenon was subsequently reconsidered \cite{shiba,bula,glma,spki},
but in a richer and physically realizable context -- tunneling through an
Anderson impurity.  When the tunneling amplitude $t$ onto the impurity
is small,
the ground state is a $\pi$-junction provided that the impurity is singly
occupied,  {\it i.e.\/}\ $U>-  \varepsilon ^{\vphantom{\dagger}}_0>0$.  We have derived the (approximate) nonperturbative
phase diagram for the junction valid at finite $t$, and in so doing have
discovered the existence of two additional phases, denoted ${\bf 0}'$ and
$ {\mib\pi}'$, in which the ground state energy $E(\delta)$ has a cusp at some
$\delta\in[0,\pi]$.  This behavior is manifested in the $I( {\bar V})$
characteristic of the junction, where for small voltage $ {\bar V}$
the current approaches the critical current $I_ {\rm c}$ extremely rapidly.
It is also clear that systems with magnetic impurities should be sensitive
to an external magnetic field.  However, tunneling experiments reported in
\cite{YN} demonstrated the absence of such dependence up to fields of 6 T.
This could be explained by an antiferromagnetic interaction between the
impurity sites, a possibility we are now investigating.

We gratefully acknowledge conversations with J. Hirsch, S. Liu,
and D. S. Rokhsar.  We are particularly grateful to S. Kivelson and
L. Glazman for several conversations, suggestions, and critical reading
of the manuscript.

\section{Appendix A}
In this appendix we discuss some subtleties associated with the
Hubbard-Stratonovich transformation for interacting fermions.
Suppose $\gamma(\tau)$ is distributed according to
\begin{equation}
P[\gamma(\tau)]= {\cal N}\,\exp\left(-{1\over 2U} \int\limits_0^\beta\!d\tau\,\gamma^2(\tau)
\right)
\end{equation}
where $ {\cal N}$ normalizes the distribution.  Then averaging with respect to
$P[\gamma(\tau)]$ one finds, for $M(\tau)$ a commuting variable,
\begin{displaymath}
\bigg\langle\exp\left(- \int\limits_0^\beta\!d\tau\,\gamma(\tau)\,M(\tau)\right)
\bigg\rangle=\exp\left({U\over 2} \int\limits_0^\beta\!d\tau\,M^2(\tau)\right)\ .
\end{displaymath}
Now consider the simplest Hamiltonian
imaginable -- a single spinless fermion with $ {\cal H}_ {\rm a}=-\mu\,c ^\dagger c$.
We compute the partition function via a coherent state path integral, and
since if we let $M(\tau)= {\bar c}(\tau)\,c(\tau)$, then from $M^2(\tau)=0$
we obtain the result
\begin{eqnarray}
Z_ {\rm a}&=&\bigg\langle \int\! {\cal D}[ {\bar c},c]\,\exp\left(- \int\limits_0^\beta\!d\tau\,
 {\bar c}\,( {\partial}_\tau+\gamma(\tau)-\mu)\,c(\tau)\right)\bigg\rangle\nonumber\\
&=&\Big\langle {\rm det}\,( {\partial}_\tau+\gamma(\tau)-\mu)\Big\rangle\nonumber\\
&=&\bigg\langle 1+\exp\left(\beta\mu- \int\limits_0^\beta\!d\tau\,\gamma(\tau)\right)
\bigg\rangle\nonumber\\
&=&1+e^{\mu/T}\,e^{U/2T}
\label{za}
\end{eqnarray}
which is incorrect -- the $e^{U/2T}$ factor should not be present in the
second term.

If we consider another trivial model,
\begin{equation}
 {\cal H}_ {\rm b}=-\mu( c ^\dagger_\uparrow c ^{\vphantom{\dagger}}_\uparrow+ c ^\dagger_\downarrow c ^{\vphantom{\dagger}}_\downarrow) + U  c ^\dagger_\uparrow c ^\dagger_\downarrow c ^{\vphantom{\dagger}}_\downarrow c ^{\vphantom{\dagger}}_\uparrow
\end{equation}
and let $M(\tau)= \bar c ^{\vphantom{\dagger}}_\uparrow  c ^{\vphantom{\dagger}}_\uparrow- \bar c ^{\vphantom{\dagger}}_\downarrow  c ^{\vphantom{\dagger}}_\downarrow$,
we again obtain an incorrect result,
\begin{equation}
Z_ {\rm b}=1+2\,e^{\mu/T}\,e^{U/2T}+ e^{2\mu/T}\ .
\label{zb}
\end{equation}

These problems do not appear in the discrete time version of the fermionic
path integral.  The discrete action is, for spinless fermions,
\begin{equation}
S_{\rm discrete}=\sum_{i,j}^N  {\bar c}_i\, M_{ij}\, c_j
\end{equation}
where $N=\beta/ \epsilon$ is the number of time slices, and, for $ {\cal H}_ {\rm a}$,
the matrix $M_{ij}$ is
\begin{displaymath}
\pmatrix{1&0& & \cdots & 1- \epsilon {\tilde\gamma} ^{\vphantom{\dagger}}_N\cr
-1+ \epsilon {\tilde\gamma} ^{\vphantom{\dagger}}_1 & 1 & 0 & & 0\cr
0 & -1+ \epsilon {\tilde\gamma}_2 & 1 & & \vdots \cr
\vdots & & & & 0 \cr
0 & \cdots & 0 & -1+ \epsilon {\tilde\gamma} ^{\vphantom{\dagger}}_{N-1} & 1\cr}
\end{displaymath}
where $ {\tilde\gamma} ^{\vphantom{\dagger}}_n\equiv \gamma ^{\vphantom{\dagger}}_n-\mu$.  We must now integrate out
the Grassmann variables and then average over the distributions
\begin{displaymath}
P(\gamma_j)=\left({ \epsilon\over 2\pi U}\right)^{1/2}\,\exp(- \epsilon\gamma_j^2/2U)
\end{displaymath}
for each time slice $j$.  But now we find
\begin{equation}
{\rm det}\,M=1+\prod_{j=1}^N (1- \epsilon\gamma_j+ \epsilon\mu)
\end{equation}
and indeed
\begin{eqnarray}
Z_ {\rm a}&=&\big\langle {\rm det}\,M (\gamma ^{\vphantom{\dagger}}_1,\ldots,\gamma ^{\vphantom{\dagger}}_N)
\big\rangle ^{\vphantom{\dagger}}_{\{\gamma ^{\vphantom{\dagger}}_i\}}\nonumber\\
&=&1+(1+ \epsilon\mu)^N\nonumber\\
&=& 1+e^{\mu/T}\ .
\end{eqnarray}

When we apply the discrete time path integral to $ {\cal H}_ {\rm b}$, we find
\begin{eqnarray}
{\rm det}\,M ^{\vphantom{\dagger}}_ \uparrow\>{\rm det}\,M ^{\vphantom{\dagger}}_ \downarrow
&=&1+\prod_{j=1}^N (1- \epsilon\gamma_j+ \epsilon\mu)+
\prod_{j=1}^N (1+ \epsilon\gamma_j+ \epsilon\mu)\nonumber\\
&&\qquad+\prod_{j=1}^N \left( (1+ \epsilon\mu)^2- \epsilon^2\gamma_j^2 \right)
\end{eqnarray}
and we find, correctly,
\begin{equation}
Z_ {\rm b}=1+2\,e^{\mu/T}+e^{(2\mu-U)/T}\ .
\end{equation}

The essence of the problem is that in the discrete case we have
\begin{displaymath}
\langle \>1- \epsilon\gamma_j\>\rangle = 1
\end{displaymath}
whereas if we write $1- \epsilon\gamma\approx \exp(- \epsilon\gamma)$ and then average,
\begin{displaymath}
\langle \exp(- \epsilon\gamma_j)\rangle = \exp( \frac{1}{2} \epsilon U)\ ,
\end{displaymath}
which gives an incorrect contribution at $ {\cal O}(\varepsilon)$.

The remedy for this frustrating problem is to perform a shift
\begin{equation}
\mu\to {\tilde\mu}\equiv \mu- \frac{1}{2} U
\end{equation}
in the continuum case.  This rescues the correct results from both
equations (\ref{za}) and (\ref{zb}), and may be formally obtained
by point splitting $M^2(\tau)\to M(\tau^+)\,M(\tau^-)$.

\section{Appendix B}
To demonstrate the reliability of the formalism and approximations used
in this paper, we consider the toy model
\begin{displaymath}
 {\cal H}_ {\rm c}=-\mu( c ^\dagger_\uparrow c ^{\vphantom{\dagger}}_\uparrow+ c ^\dagger_\downarrow c ^{\vphantom{\dagger}}_\downarrow)+
\Delta( c ^\dagger_\uparrow c ^\dagger_\downarrow+ c ^{\vphantom{\dagger}}_\downarrow c ^{\vphantom{\dagger}}_\uparrow)+ U  c ^\dagger_\uparrow c ^\dagger_\downarrow c ^{\vphantom{\dagger}}_\downarrow c ^{\vphantom{\dagger}}_\uparrow\ .
\end{displaymath}
Once again we break up the interaction term with a Hubbard-Stratonovich
transformation, careful to substitute $\mu\to {\tilde\mu}$.  All the time dependence
of the field $\gamma(\tau)$ can be gauged away via the transformation
\begin{eqnarray}
c(\tau)&\to& e^{-g(\tau)}\,c(\tau)\nonumber\\
 {\partial}_\tau g(\tau)&=&\gamma(\tau)-{1\over\beta} \int\limits_0^\beta\!d\tau'\,
\gamma(\tau')\ .
\end{eqnarray}
Note that $g(\beta)=g(0)$, so the antiperiodic boundary conditions
$c(\beta)=-c(0)$ are not affected by this transformation.  Thus,
\begin{eqnarray}
Z_ {\rm c}&=&{1\over\sqrt{2\pi U T}} \int\limits_{-\infty}^\infty \!d\gamma\,
e^{-\gamma^2/2UT}\,e^{( {\tilde\mu}+\gamma)/T}\nonumber\\
&&\qquad\times\, {\rm det}\,
\pmatrix{ {\partial}_\tau+\gamma- {\tilde\mu} &\Delta\cr\Delta & {\partial}_\tau+\gamma+ {\tilde\mu}\cr}\ ,
\end{eqnarray}
where now $\gamma$ is time-independent, denoting simply the average
of what we have until now been calling the function $\gamma(\tau)$.
Now if $\{\lambda_0,\ldots,\lambda_3\}$ are constants, then
\begin{equation}
{\rm det}\,( {\partial}_\tau+\lambda_0 +  {\vec\lambda}\cdot {\vec\sigma})
=1+2\,e^{-\lambda_0/T}\,\cosh{| {\vec\lambda}|\over T} + e^{-2\lambda_0/T}\ .
\end{equation}
Performing the integral over $\gamma$, one correctly obtains the partition
function which may be calculated by more elementary means.

\begin{figure} [!t]
\centering
\leavevmode
\epsfxsize=8cm
\epsfysize=8cm
\epsfbox[18 144 592 718] {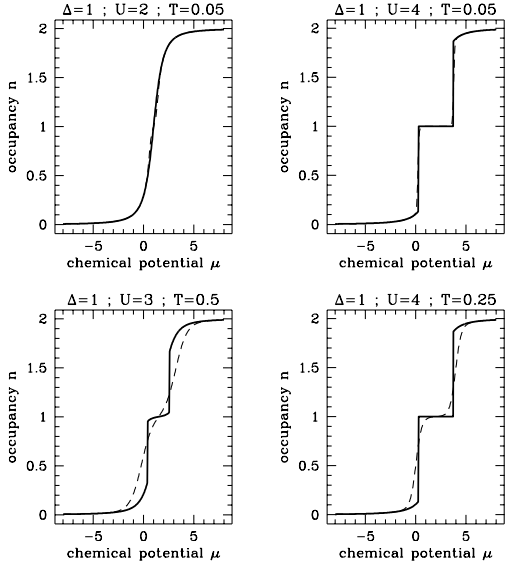}
\caption[]
{\label{toy} Occupancy $n$ {\it versus\/} bare chemical potential $\mu$
for the toy model $ {\cal H}_ {\rm c}$ computed in the steepest descents
approximation (solid) and compared with exact results (dashed).}
\end{figure}

The steepest descents approximation (SDA) to the integral over $\gamma$ gives
\begin{equation}
{\gamma\over U}={1-e^{-2\gamma/T}\over
1+2\,e^{-\gamma/T}\,\cosh{ \sqrt{{\tilde\mu}^2+\Delta^2}\over T}+e^{-2\gamma/T}}\ ,
\label{mfeqn}
\end{equation}
which is obtained by extremizing
\begin{eqnarray}
F&=&{\gamma^2\over 2U}-\gamma- {\tilde\mu}
-T\ln\left(1+e^{-\left(\gamma- \sqrt{{\tilde\mu}^2+\Delta^2}\right)/T}\right)\nonumber\\
&&\qquad-T\ln\left(1+e^{-\left(\gamma+ \sqrt{{\tilde\mu}^2+\Delta^2}\right)/T}\right)
\end{eqnarray}
with respect to $\gamma$.  $\gamma=0$ is always a solution, and for
a range of parameters it is possible to have more than one solution
to (\ref{mfeqn}), in which case we choose the solution with the lowest
free energy.  Once a solution for $\gamma$ is obtained,
we evaluate the particle number
\begin{eqnarray}
N&=&-{ {\partial} F\over {\partial} {\tilde\mu}}\\
&=&1+{ {\tilde\mu}\over \sqrt{{\tilde\mu}^2+\Delta^2}}\,
{2\,e^{-\gamma/T}\,\sinh{ \sqrt{{\tilde\mu}^2+\Delta^2}\over T}\over
1+2\,e^{-\gamma/T}\,\cosh{ \sqrt{{\tilde\mu}^2+\Delta^2}\over T}+e^{-2\gamma/T}}\ .\nonumber
\end{eqnarray}
Comparisons of the SDA with the exact results are shown in
FIG. \ref{toy}.  For the Hamiltonian $ {\cal H}_ {\rm c}$, the SDA is exact
in the limit of zero temperatures.  At finite $T$, the SDA leads to
isolated discontinuous changes of $\gamma$ as $\mu$, $U$, and $\Delta$
are varied, which are reflected in the behavior of  {\it e.g.\/}\ $n(\mu)$.
The exact solution behaves smoothly, but for sufficiently low temperatures the
agreement is arbitrarily good.


\section{Bibliography}

\begin{references}

\bibitem{kulik} I. O. Kulik, {\sl Sov. Physics JETP}, {\bf 22}, 841 (1966).

\bibitem{shiba} H. Shiba and T. Soda, {\sl Prog. Theor. Phys.} {\bf 41}, 25
(1969).

\bibitem{glma} L. I. Glazman and K. A. Matveev, {\sl Pis'ma Zh. Eksp. Teor.
Fiz.} {\bf 49}, 570 (1989) [{\sl JETP Lett.} {\bf 49}, 659 (1989)].

\bibitem{spki} B. I. Spivak and S. A. Kivelson, {\sl Phys. Rev. B}
{\bf 43}, 3740 (1991).

\bibitem{bula} L. N. Bulaevskii, V. V. Kuzii, and A. A. Sobyanin,
{\sl JETP Lett.} {\bf 25}, 290 (1977)

\bibitem{Harlingen} D. J. Van Harlingen, {\sl Rev. Mod. Phys.} {\bf 67},
515 (1995).

\bibitem{Kondo} I. Affleck and A. W. W. Ludwig, {\sl Nucl. Phys. B}
{\bf 330}, 641 (1991).

\bibitem{ChamFrad} C. de C. Chamon and E. Fradkin, {\sl Phys. Rev. B}
{\bf 56}, 2012 (1997).

\bibitem{foot1} The factor of $ k_{\scriptscriptstyle{\rm F}}^{-1/2}$ multiplying $t$ is inserted
so that $t$ may have the dimensions of energy.

\bibitem{TanKash} Y. Tanaka and S. Kashiwaya, {\sl Physica C} {\bf 274},
357 (1997).

\bibitem{lgpc} L. Glazman, private communications.

\bibitem{YN} J. Yoshida and T. Nagano, {\sl Phys. Rev. B}, {\bf 55}, 11860
(1997).



\end{references}
\end{document}